# Interface stability of β-Ga$_2$O$_3$ (100) on oxidized Si- and C-terminated 3C-SiC (001) substrates: a first-principles investigation

M. Licciardi[a], A. Ugolotti[a*], E. Scalise[a] and L. Miglio[a]

[a]Department of Materials Science, University of Milano-Bicocca, via Roberto Cozzi, 55, 20126 Milano (MI), Italy

* email: aldo.ugolotti@unimib.it

**ABSTRACT**

We provide a first-principles modeling of the β-Ga$_2$O$_3$/3C-SiC interface that takes into account the reconstructions occurring at the 3C-SiC (001) surface by oxidation, aiming to mimic the actual deposition process under the best structural and thermodynamic conditions. Using density functional theory calculations, we systematically investigate the interface configurations between β-Ga$_2$O$_3$ (100) and both Si- and C-terminated 3C-SiC (001) substrates, considering realistic oxidation states that form at the SiC surface prior to epitaxial growth. Our analysis evaluates different stacking sequences and atomic-scale bonding arrangements, computing adhesion energies for various interface geometries to determine their relative thermodynamic stability. This work addresses the critical need for understanding β-Ga$_2$O$_3$ integration on substrates with superior thermal conductivity, providing a theoretical framework for optimizing heteroepitaxial growth conditions in ultra-wide-bandgap power electronics applications.

## 1) INTRODUCTION

4H-SiC and GaN have emerged as suitable materials for power electronics [1,2], showing superior properties compared to silicon [3,4]. However, they still lack a silicon-integrated, scalable, and low-cost fabrication method. Additionally, their bandgap values -respectively 3.26 eV and 3.39 eV [3]- are too small to handle high voltages, as for renewable energies and industrial power supplies. For such applications, promising materials are Ultra-Wide-Band-Gap (UWBG) semiconductors [4,5], characterized by even larger bandgap values. Among UWBG oxides [5,6], the most studied one is gallium oxide (Ga$_2$O$_3$) [7–11], in particular its thermodynamically stable phase β-Ga$_2$O$_3$, which can be obtained in bulk crystals by low-cost melt-methods [7].

β-Ga$_2$O$_3$ exhibits a direct bandgap of approximately 4.9 eV [12,13] that leads to an exceptionally high breakdown field [4], making β-Ga$_2$O$_3$ an optimal candidate for high voltage and high power (> 1kW) applications [10]. The poor thermal conductivity of β-Ga$_2$O$_3$ (~30 W/mK) necessitates the development of thermal management strategies, such as integrating it into materials with significantly higher thermal conductivity, like SiC. Integration with silicon would also be very interesting for smart power devices, driven by logic transistors or sensors. For these reasons, epitaxy techniques, like Molecular Beam Epitaxy (MBE) and Metal-Organic Chemical Vapour Deposition (MOCVD) [10,14–17], are widely employed in the deposition of β-Ga$_2$O. To date, β-Ga$_2$O$_3$ heteroepitaxial growth has been predominantly realized on c-plane sapphire (Al$_2$O$_3$) substrates [15] due to their widespread availability and relatively low cost. In addition, the oxygen-terminated surface can accommodate a continuous oxygen network common to both film and substrate. This simplifies the initial nucleation and often promotes the formation of well-oriented β-Ga$_2$O$_3$ epitaxial layers, using the already mentioned techniques. However, the lattice mismatch between β-Ga$_2$O$_3$ and sapphire -9.5% and 3.0% in the in-plane directions- inevitably introduces significant strain in the film, possibly affecting phase stability, rotational domain formation, and overall crystal quality. We have already extensively discussed the strain effects on phase competition and interface energetics in previous works [18,19]. Actually, there is growing interest in identifying alternative substrates that offer close lattice symmetry and lower misfit strain, therefore supporting high-quality epitaxy. In particular, an ongoing challenge in the power electronics industry is the efficient integration of PDs with silicon-based technology, in order to leverage the maturity of silicon wafer fabrication -improving the device scalability and reducing the costs- and to enable on-chip gate logic control [20]. For what concerns β-Ga$_2$O$_3$, the heteroepitaxy on silicon substrate is quite challenging [21], because of structural and thermal misfits between the two materials. Additionally, the concurrent deep oxidation of silicon represents a further obstacle towards Ga$_2$O$_3$ deposition on Si. Then, the insertion of a buffer layer is a viable path for integrating β-Ga$_2$O$_3$ on silicon substrates [20,22]. A good choice as the buffer layer material is the cubic phase of silicon carbide (3C-SiC) [23,24]. It is inert to deep oxidation and has a very good thermal conductivity [25], compensating for the poor one of β-Ga$_2$O$_3$. Its lattice parameters are smaller compared to silicon ones, thus the structural misfit between β-Ga$_2$O$_3$ and silicon is mitigated between the two adjoining layers. Lastly, the deposition of 3C-SiC on silicon is already an optimized process, making it necessary to investigate only the oxide growth on 3C-SiC / Si substrates.

At present, there are few studies about β-Ga$_2$O$_3$ heteroepitaxy on 3C-SiC. Kukushkin et al. [21] in 2016 reported a continuous film of β-Ga$_2$O$_3$ grown by hydride-chloride vapor phase epitaxy (HVPE) on a 3C-SiC layer deposited on top of a Si (111) substrate by chemical substitution [26]. The sample characterization showed that the β-Ga$_2$O$_3$ film is close to the epitaxial structure with a (201) orientation and a polycrystalline phase is absent. This was the first demonstration of the feasibility of growing β-Ga$_2$O$_3$ on silicon thanks to a 3C-SiC buffer layer. Lately, Osipov et al. [27] used the same technique to grow β-Ga$_2$O$_3$ films on silicon substrates with different crystallographic orientations –[100], [110], and [111]-, to take into account the anisotropy of crystalline materials.

The computational work presented here aims at investigating in detail the possible β-Ga$_2$O$_3$/3C-SiC interfaces, setting the theoretical foundations to predict which epitaxial structure is more preferably obtained under experimental conditions. Among the different crystallographic orientations of 3C-SiC grown on silicon, the (111) surface should be discarded, because its hexagonal symmetry promotes the formation of C$_3$ rotational domains in the oxide film [28,29], leading to rotational disorder and defect structures that are undesirable for device applications. Between the other two possible crystallographic orientations, namely (001) and (110), it is necessary to determine which one better matches the β-Ga₂O₃ lattice. In fact, a lower mismatch usually leads to a higher quality film, reducing the probability of defect formation. Additionally, it is important to mention that Boschi et al. [30] experimentally observed the spontaneous formation of a thin β-Ga₂O₃ layer on 3C-SiC (001), even under conditions optimized for the ε-phase. This result could suggest that β-Ga₂O₃ growth on 3C-SiC (001) might be favored.

In this work, we first determine the best match between the substrate and the film in terms of surface misfits. Then, for the selected 3C-SiC surface, we investigate in detail its reconstruction and the early stages of oxidation, in order to clarify how the substrate behaves when exposed to the oxygen flux and if it is fully inert to oxidation. Finally, we construct and analyze several possible β-Ga₂O₃ / 3C-SiC interface configurations, evaluating their energies to determine the most favorable one and to assess the suitability of 3C-SiC as a substrate for the heteroepitaxial growth of β-Ga

## 2) COMPUTATIONAL METHODS

All our calculations were carried out in the framework of the density functional theory (DFT) by the VASP software [31–33], using the Perdew–Burke–Ernzerhof exchange-correlation functional revised for solids (PBEsol) [34] and GW pseudopotentials. In addition, for Ga atoms, the d-shell was also included in the valence states. The lattice parameters of 3C-SiC were optimized using a plane-waves cutoff of 600 eV and sampling the reciprocal lattice with a mesh of 4x4x4 k-points centered in Γ point. We find a lattice constant a=b=c=4.3585 Å (3C-SiC), in good agreement with both computational and experimental results in the literature [35,36]. The ($\sqrt{2}x\sqrt{2}$)R45 supercell along the a,b plane defines the surface p(2x2) structure that we will use in the following to construct our substrate geometry.

The bulk of $\beta$-Ga₂O₃ was first oriented to align its (100) surface with the (001) surface of the 3C-SiC crystal and we matched the a,b lattice parameters of such an oriented cell with those of the p(2x1) cell of the 3C-SiC. Then, the strained $\beta$-Ga₂O₃ bulk was optimized, using a plane-waves cutoff of 800 eV and a 12x6x3 mesh of k-points, to consistently obtain the cell height and the atomic coordinates.

We studied both surfaces and interfaces in a slab geometry, including a vacuum region of at least 13 Å and a dipole correction along the z axis, to decouple the periodic images; all these slab calculations were carried out with a plane-waves cutoff of 500 eV.

We built the 3C-SiC (001) surface as one (2x2) supercell of the p(1x1) surface lattice, to allow the surface reconstruction, with ten SiC bilayers. We mapped the reciprocal space with a 3x3x1 k-points mesh. For such slabs, we expose symmetric terminations, either Si- or C-terminated. Since the resulting structures are non-stoichiometric, the surface energies were determined as a function of the chemical potential of Si:

$$\gamma_{surf}^{SiC}(\Delta\mu_{Si}) = \frac{E_{slab} - N_{SiC}\,\mu_{SiC} - N_{Si}\,\mu_{Si}}{2\,A} \quad (1)$$

where $E_{slab}$ is the total energy of the surface slab, $\mu_{SiC}$ is the chemical potential of the bulk 3C-SiC, and $N_{SiC}$ is the number of SiC units in the slab. $N_{Si}$ accounts for the number of non-stoichiometric Si atoms, which could be positive or negative, and A is the superficial area of the slab. The values of $\mu_{Si}$ must be limited to those allowing a state of equilibrium for the solid phase: Si-bulk limit, where $\mu_{Si}$ equals the Si bulk value (i.e., $\Delta\mu_{Si} = \mu_{Si} - \mu_{Si}^{bulk} = 0$) and C-bulk limit, where $\mu_C$ is fixed to the value calculated for the bulk diamond phase. Thus, $\Delta\mu_{Si}$ is obtained imposing $\Delta\mu_{Si} = \mu_{3C-SiC} - \mu_C^{diamond} - \mu_{Si}^{bulk} = \Delta H_{SiC}$ (i.e., heat of formation of SiC).

In the study of oxidation, of both Si- and C-surface terminations, since we are dealing with adsorbates, the Grimme-D3 [37] Van der Waals correction was included in our calculations. To establish the most favored adsorption sites, we calculated the adsorption energy of N oxygen atoms as:

$$E_{ads}^{(N)} = E_{tot}^{(N)} - E_{slab}^{pristine} - \frac{N}{2}\mu_{O_2} \quad (2)$$

where $E_{tot}^{(N)}$ and $E_{slab}^{pristine}$ are the total energy of the slab with N oxygen atoms and no adatoms, respectively and $\mu_{O_2}$ is the chemical potential calculated for an isolated oxygen molecule. Finally, we simulated the β-Ga$_2$O$_3$(100)/3C-SiC (001) heterostructures by means of slabs formed by ten Si-C bilayers, four Ga-O bilayers and a thin layer of a variable number of non-stoichiometric oxygen atoms in between. Structural relaxation was performed by optimizing only the atomic coordinates while keeping the supercell parameters fixed, ensuring proper stress release along the out-of-plane direction only. Then, the interface energies have been calculated as:

$$\gamma_{int}(\mu_O, \mu_{Si}) = \frac{E_{slab} - N_{SiC}\,\mu_{SiC} - N_{Ga_2O_3}\,\mu_{Ga_2O_3} - A\gamma_{surf}^{SiC}(\mu_{Si}) - A\gamma_{surf}^{Ga_2O_3} - N_O\,\mu_O}{A} \quad (3)$$

where $N_O$ is the number of non-stoichiometric oxygen atoms and $\mu_O$ their chemical potential. We let the latter span between O-rich conditions, i.e. the chemical potential of molecular oxygen ($\Delta\mu_O = \mu_O - \mu_{O_2}/2 = 0$), and Ga-rich conditions, i.e. when $\Delta\mu_O = \mu_O^{Ga-rich} - \mu_{O_2}/2 = (\mu_{Ga2O3} - 2*\mu_{Ga}^{bulk})/3 - \mu_{O_2}/2$. For these structures, we relied on the grid of k-points of Ga$_2$O$_3$, properly rescaled to maintain in the reciprocal space of the interfaces' supercell the density of k-points defined for the bulk.

## 3) RESULTS AND DISCUSSION

In order to construct reasonable guesses for the structure of the interfaces of β-Ga$_2$O$_3$ with 3C-SiC, the first step of our study is to define the most promising combination of phase and orientation of the film with the substrate. Therefore, we perform a geometrical analysis evaluating both the misfit strain on the film and the epitaxial matrix of the match of the surface lattices of the two materials. The results of this analysis are collected in Table S1. Details on the different Ga$_2$O$_3$ surfaces are reported elsewhere [18]. The comparison of these preliminary results shows that several interfaces are characterized by relatively low values of the misfit strain, i.e. ε< 2%, but all of them have only non-unitary epitaxial matrices. This would imply the presence of multiple dangling bonds and, consequently, of interface defects. The notable exception is the β-$Ga_2O_3$ (100) surface on reconstructed 3C-SiC (001), for which there is a direct one-to-one matching with the substrate, at the price of a larger lattice mismatch (~6.0% along the [001] direction, ~1.1% along the [010] one). Moreover, β-$Ga_2O_3$ (100) surface is known to be particularly low in energy [18,38] due to the few dangling bonds per surface unit cell. Therefore, we will focus on 3C-SiC (001) and β-$Ga_2O_3$ (100) surfaces to construct our interfaces.

### *3.1) 3C-SiC (001) surface: relative stability and oxidation*

We first examine in detail the ideal Si- and C-terminated 3C-SiC (001) surfaces [23,39–41]. For both 3C-SiC terminations, different reconstructions have already been reported in the literature [40,41], since the Si and C atoms at the surface tend to form dimers to saturate their dangling bonds after cleavage. Regarding the Si-terminated surface, we focus on the p(2x1) [42] (Figure 1), occurring at strain-free surfaces with little contamination [42,43]. In the case of the C-terminated surface, we consider three models currently debated in the literature, namely $C_2$ bridging groups (B reconstruction), row, and staggered C-dimers (RD and SD, respectively, shown in Figure 1). For all the reconstructions investigated, we calculate the surface energy as expressed in Equation 1. The Si-terminated surface energy is in the range of 0.17-0.20 eV/Å² for (001), while the C-terminated reconstructions have slightly higher values, 0.22-0.26 eV/Å², due to the strong rearrangement to form the C-dimers. Although our calculations indicate the RD reconstruction as the most stable, experimental observations consistently report the B reconstruction [44,45]. Previous studies have shown that the choice of the computational parameters could influence the order of the relative stability of these reconstructions [46], since they are closely competing. We therefore adopt the experimentally observed B reconstruction as the reference surface for investigating oxidation processes and the subsequent deposition of β-$Ga_2O_3$.

*Si-terminated oxidation*. On the Si-terminated p(2×1) surface reconstruction, we identify four equivalent adsorption bridge sites in our cell, which were occupied progressively, also accounting for different filling orders. The whole set of optimized configurations is reported in Figure S1; since the two rows in the supercell are almost equivalent, we extrapolate two models, which are shown in Figure 2b and 2c, corresponding to half and full coverage of O adatoms (in the same row). In all cases, oxidation is highly favored since each Si dangling bond is passivated by oxygen. This is also supported by interatomic distances. The reconstructed surface shows a significantly longer Si-Si distance compared to the pristine Si (001) reconstruction, as shown in Figure 2a. On the contrary, with an O adatom, the Si-O and Si-C distances are comparable to those found in the bulk. Moreover, the energy gain of the adsorption is similar in the two cases, in the range of -5 to -6 eV, indicating that the two sites are rather independent.

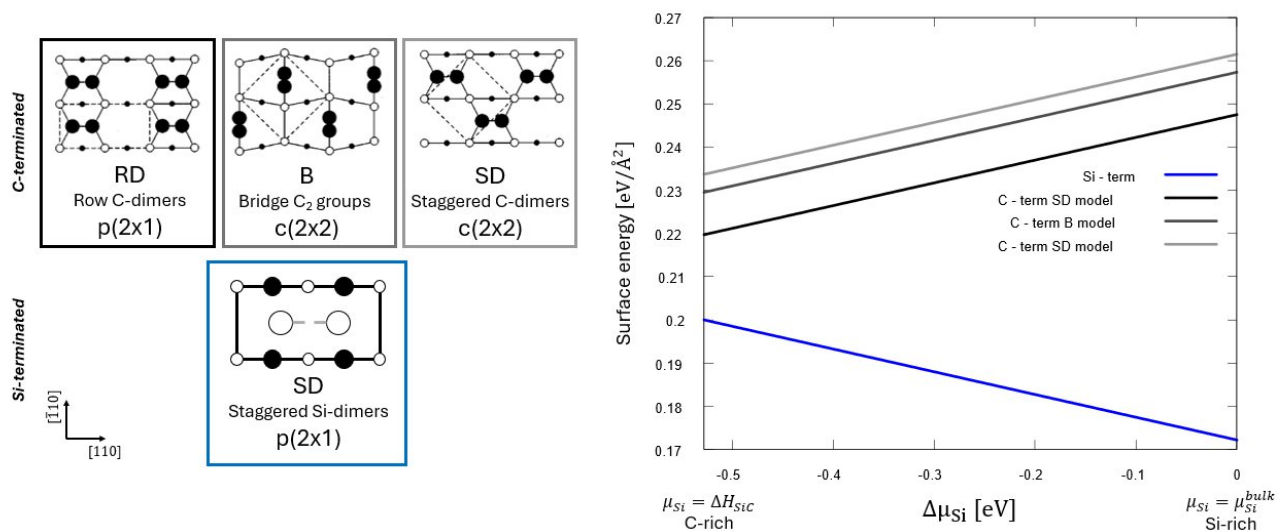

**Figure 1** – Surface energies and model reconstructions of the p(2x1) Si-termination and the row dimer (RD), bridge $C_2$ (B) and staggered dimer (SD) reconstructions of the C-termination of 3C-SiC (001). The surface energy is expressed as a function of the silicon chemical potential difference $\Delta\mu_{Si}$.

We also include additional oxidation steps, up to eight oxygen atoms per simulation cell, corresponding to the further oxidation of an already completed monolayer, as reported in Figure S2. We highlight that reaching the total of eight O atoms per cell, the structure spontaneously optimizes into a $SiO_2$ monolayer, just weakly interacting with a C-terminated 3C-SiC slab.

*C-terminated oxidation*. In the case of the C-terminated c(2×2) 3C-SiC (001) surface, we focus on the bridge $C_2$ reconstruction, as it is the most frequently observed structure experimentally [41,45]. For this surface, the analysis of the oxidation pattern is more complex, due to the presence of multiple inequivalent adsorption sites, as shown in Figure 3a. Given the complexity of this termination, we've focused on a specific site, particularly the A site, as shown in Figure 3b, among all the investigated adsorption configurations (Figure S3). While this isn't the most favored adsorption site, its unique pillar-like structure is likely to significantly influence $Ga_2O_3$ film epitaxy. Thus, this configuration, as discussed below, is used to construct the C-terminated interfaces.

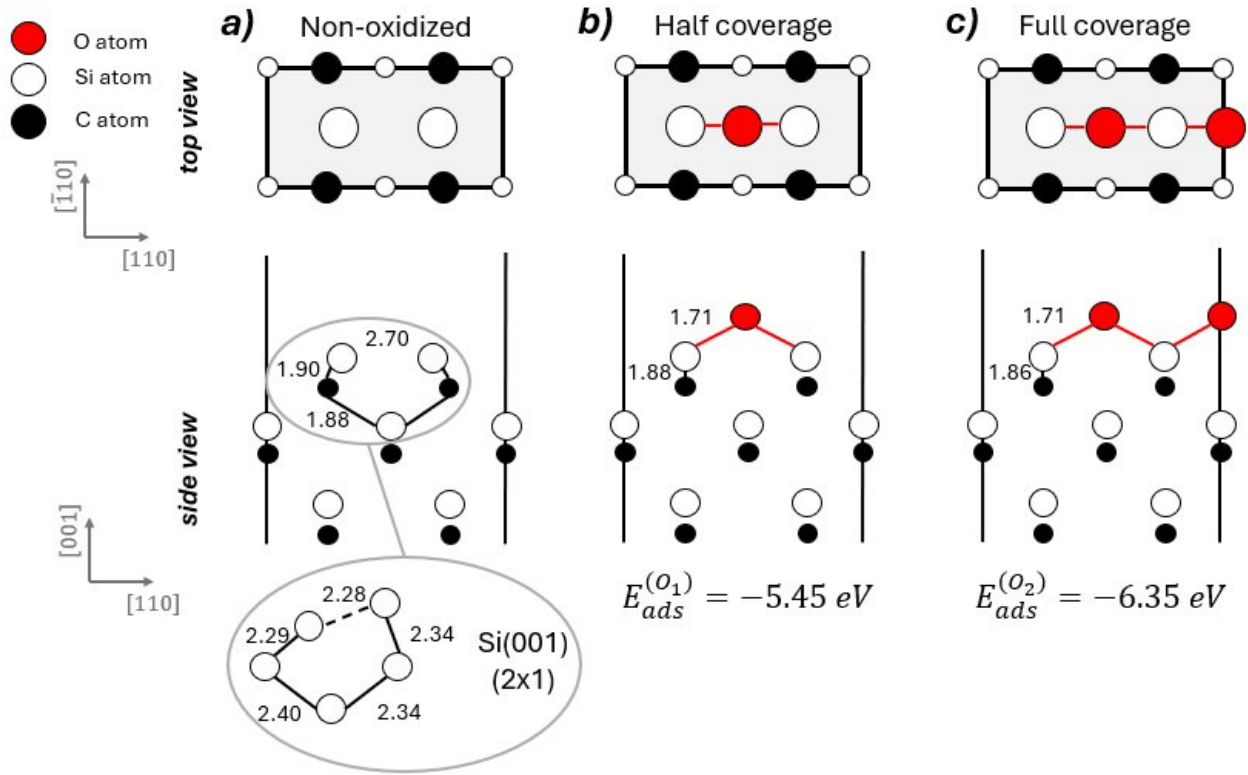

**Figure 2**: Top and side views of the p(2x1) Si-terminated 3C-SiC (001) surfaces: non-oxidized (panel a), half (panel b) and full (panel c) oxidation coverage. The relevant distances are reported in Å and the pristine Si (001) reconstruction is reported for comparison.

3.2 *Interfaces*

Before constructing the interface geometries, we shortly studied the β-$Ga_2O_3$ (100) surface. For this orientation, two possible terminations are possible, shown in Figure S4, for which we calculated a (strained) surface energy of 0.043 eV/Å$^2$ and 0.066 eV/Å$^2$, consistent with the values reported in the literature for the strain-free surfaces [18].

*Si-terminated interfaces.* We built the β-$Ga_2O_3$ (100)/Si-terminated-3C-SiC (001) interfaces by considering the half and full oxygen coverage in the p(2x1) Si reconstruction (Figure 2). We model the bottom surface of the 3C-SiC slab using the row dimer model of the C-termination, which has the same periodicity as the top Si-terminated surface and effectively minimizes any spurious charge transfer from the interface by passivating the dangling bonds at the surface. For the oxide slab, both A and B inequivalent terminations of the β-$Ga_2O_3$ (100) surface are considered in symmetric geometries. At the interface, the oxygen atoms arising from the 3C-SiC oxidation are initially positioned with twofold or threefold coordination, consistent with oxygen coordination in bulk β-$Ga_2O_3$. By systematically combining substrate oxidation coverage, oxide termination, and initial oxygen coordination, a set of eight Si-terminated interface configurations is generated, as reported in Figure S5. However, after optimization, only five of them result in being stable, while in the others the two materials moved apart (Figure S6). However, it's important to highlight that the two interfaces with full coverage and B termination, but different initial oxygen coordination, become identical after optimization. All the stable geometries exhibit well-ordered, atomically sharp interfaces (Fig. 4a-d). The atoms at the junction remain close to their bulk configurations, and the oxidation proved particularly effective at saturating interfacial dangling bonds.

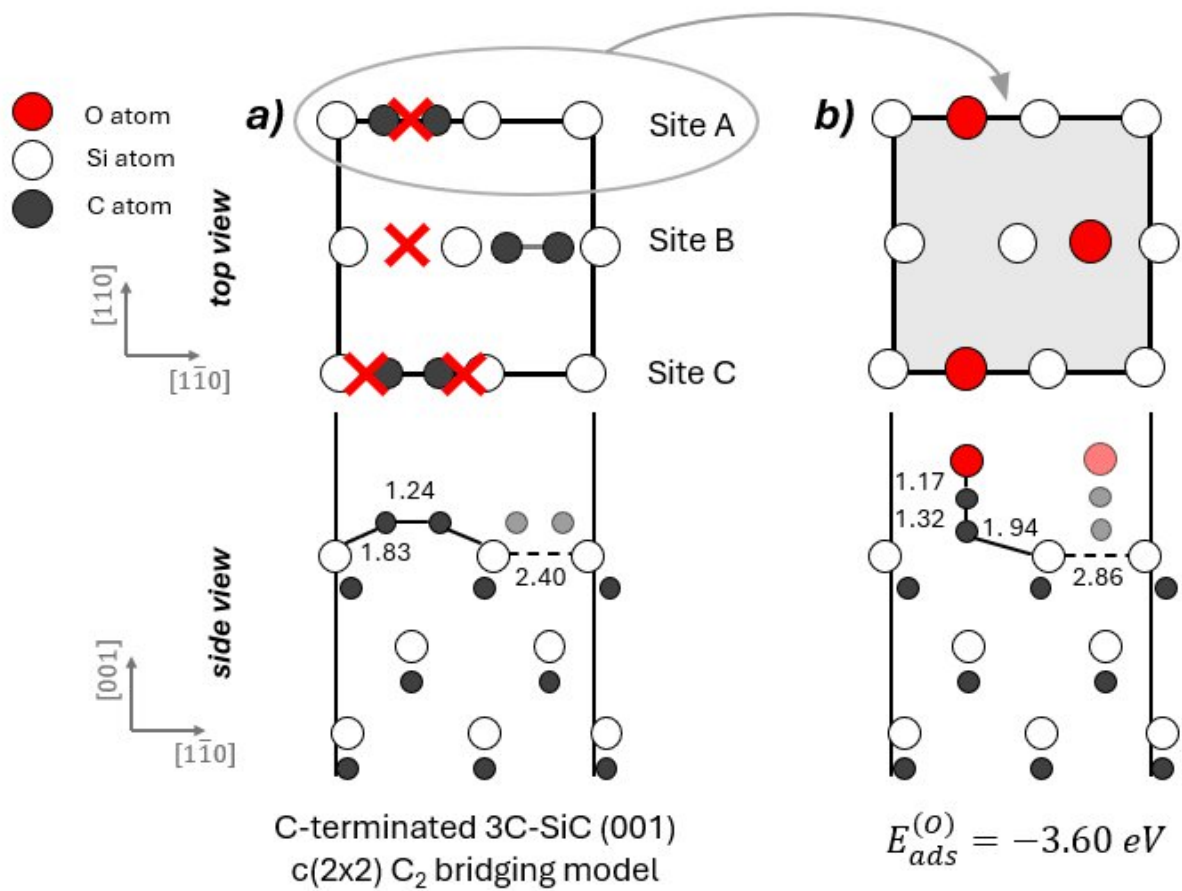

**Figure 3**: a) Top and side view of the reconstructed c(2×2) C-terminated 3C-SiC(001) surface, where the three inequivalent oxygen adsorption sites (A, B, and C) identified on the $C_2$-bridging model are marked by red crosses. b) Top and side views of the optimized structure with an O adatom at site A. The corresponding oxygen adsorption energy and interatomic distances (in Å) are reported.

To compare the thermodynamics of the different interface structures, we calculate their corresponding interface energies $\gamma_{int}$. In order to remove the degree of freedom given by the dependency of $\gamma_{int}$ on the Si chemical potential, which is just a numerical issue, we introduce the relative interface energy difference $\Delta\tilde{\gamma}_{int}$, which we define as the difference between any $\gamma^i_{int}(\mu_O)$ and the interface energy of a chosen interface. In this case, we select IS3 as a reference (Fig. 4c), since its energy does not depend on the chemical potential of oxygen, being fully stoichiometric. The plot of the $\Delta\tilde{\gamma}_{int}$ for all the interface configurations is shown in Figure 4e. The IS1 interface has an energy 0.250 eV/Å² lower than all the others at oxygen-rich conditions, which are reported in the literature as the optimal growth conditions for β-Ga$_2$O$_3$ [47]. Instead, towards Ga-rich conditions, the most stable configuration is IS3, i.e., the stoichiometric one. This can be explained by the fact that at low $\mu_O$ values, the incorporation of O atoms into the oxide structure is not favored.

Regarding the absolute interface energy of the IS1 interface, a brief comment is needed. We calculate values as low as –450 meV/Å² at O-rich conditions, indicating that interface formation is thermodynamically favored in this regime. It is fundamental to highlight that this interface energy includes an oxidation stabilization arising from the non-stoichiometric oxygens, thus obtained through oxidation. Configurations with lower oxidation coverage are still stable but exhibit progressively higher interface energies due to the partial exposure of dangling bonds that are not fully compensated by interfacial rearrangements.

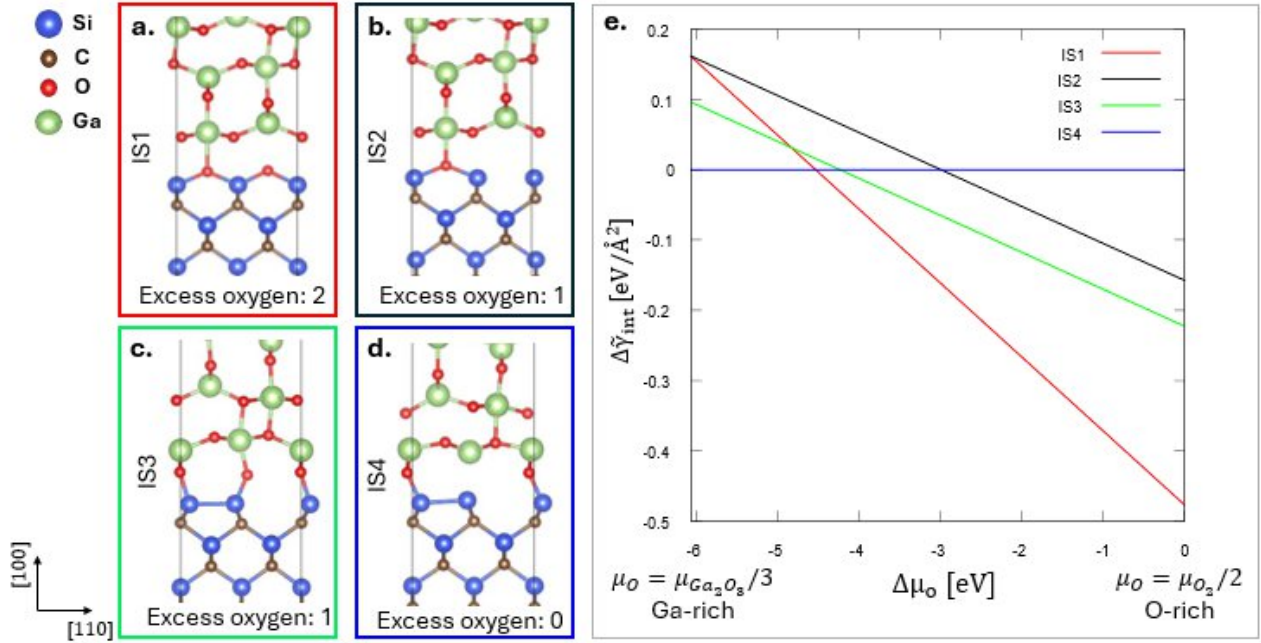

**Figure 4** – Side view of the four optimized β- Ga$_2$O$_3$ (100)/Si-terminated 3C-SiC (001) interfaces (a-d) and their relative stability evaluated in terms of $\Delta\tilde{\gamma}_{int}$ as a function of the oxygen chemical potential Δ$\mu_o$ (e).

*C-terminated interfaces*. Moving to the interface of β-Ga$_2$O$_3$ (100) with C-terminated 3C-SiC (001), we model them following a different strategy, due to the complexity of the oxidized C-terminated SiC surface. In this case, the 3C-SiC slab has p(2x2) surface periodicity and bridge C$_2$ dimers reconstruction for both the exposed surfaces. As already discussed, despite several inequivalent adsorption sites being possible, we limit our investigation to the A site (Fig. 3), since this is the configuration imposing stronger geometrical constraints. The general approach used for modelling these interfaces can be summarized as follows. We selectively replace the oxygen from oxidation with oxygen atoms in inequivalent planes of β-Ga$_2$O$_3$ (100) as shown in Figure 5a. This is followed by removing β-Ga$_2$O$_3$ atoms that are located too close to the 3C-SiC substrate, particularly those within 1.2 Å. This second step is required because while our models are constructed by joining together the different films, we aim to mimic a deposition process; accordingly, β-Ga$_2$O$_3$ atoms are allowed to occupy only positions not already occupied by 3C-SiC atoms. First, carbon pillars are replaced by oxygen atoms in the purple oxygen plane, removing all β-Ga$_2$O$_3$ atoms at short distances from 3C-SiC atoms, resulting in the first stable interface configuration (IC1, Fig. 5b). Testing the substitution of oxygen atoms from the yellow plane yields a geometry equivalent to the purple plane after removing excess atoms. Substituting carbon pillars to oxygen atoms in the grey plane and removing excessively short interatomic distances results in a distinct interfacial arrangement (structure IC2, Fig. 5c). Local bonding analysis reveals over-coordinated interfacial atoms, which are removed to obtain structure IC3 (Fig. 5d). An alternative strategy to reduce over-coordination without changing the slab composition involves slightly adjusting the positions of atoms at the interface, particularly Ga atoms, to restore coordination environments closer to those in bulk β-Ga$_2$O$_3$, resulting in structure IC4 (Fig. 5e).

In addition to the oxidized C-terminated models, we also considered possible interface configurations constructed starting from the non-oxidized substrate, namely the pristine reconstructed C-terminated 3C-SiC (001) surface. In this case, two distinct interface geometries were obtained depending on the termination of the β-Ga$_2$O$_3$ (100) slab at the junction. Specifically, interfaces were built by combining the pristine C-terminated surface with either the A (IC5, Fig. 5f) or the B (IC6, Fig.5 g) termination of β-Ga$_2$O$_3$.

Following optimization, all the heterostructures become stable and there's no separation between the two materials. The final geometries are more complex than the Si-terminated ones, especially the ones built from the oxidized surface. The interface is not abrupt and sharp, but the two materials are interlocked. These results are consistent with the predicted non-flat morphology of the oxidized C-terminated 3C-SiC (001) surface.

As before, the relative stability of the different configurations is evaluated in terms of the interface energy difference. Then, similarly to the case of the Si-terminated interfaces, we calculated $\Delta\tilde{\gamma}_{int}$ as the difference between the interface energy and that of a reference, here the IC5 (Fig. 5f).

When examining $\Delta\tilde{\gamma}_{int}$ for the C-terminated interfaces (Fig. 5h), it becomes evident that oxidation also plays a role in this case, particularly under O-rich conditions. The stabilizing effect of oxidation, however, is considerably weaker than for the Si-terminated surfaces. In C-terminated systems, the apparent stabilization at high $\Delta\mu_O$ primarily reflects the thermodynamic tendency of the system to accommodate additional oxygen atoms when the chemical potential is sufficiently large. Indeed, for $\Delta\mu_O$ values below -2.5 eV, the most stable geometries are those without non-stoichiometric oxygen, namely the interfaces constructed from the pristine, non-oxidized C-terminated surface. This behaviour can be rationalized by the fact that oxidized C-terminated interfaces undergo substantial atomic rearrangements, which carry a significant energetic cost. In contrast, interfaces derived from the pristine surface retain a relatively flat geometry where the oxide can bind without requiring major displacements of the 3C-SiC or β-Ga$_2$O$_3$ atoms from their bulk positions. Consequently, the energetic penalty of oxidation and reconstruction surpasses the modest stabilizing effect of additional oxygen under Ga-rich and intermediate conditions. This makes non-oxidized C-terminated interfaces the preferred structures across a broad range of chemical potentials. Specifically, the most stable structure is the B termination of β-Ga$_2$O$_3$ grown on a pristine C-terminated substrate.

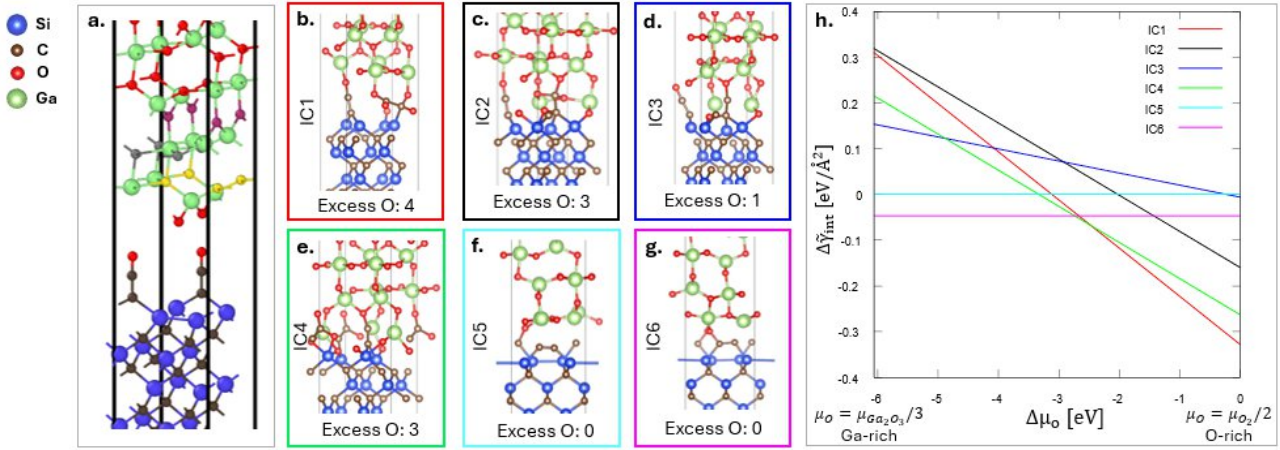

**Figure 5** - Modeling of β-Ga$_2$O$_3$ (100)/C-terminated 3C-SiC (001) heterostructure, by merging the oxidized or pristine C-terminated 3C-SiC (001) surface with the β-Ga$_2$O$_3$ (100) slab. The oxygen atoms forming vertical *pillars* at the interface are assumed to interlock with the oxide lattice by substituting oxygen atoms belonging to one of the three nonequivalent planes in β-Ga$_2$O$_3$ highlighted in green, blue, and pink (a), giving rise to different possible interface configurations (b-e). From the pristine non-oxidized C-terminated surface, two configurations are possible (f,g), varying the β-Ga$_2$O$_3$ (100) terminations. The stability of all these configurations has been evaluated in terms of $\Delta\tilde{\gamma}_{int}$ as a function of the oxygen chemical potential $\Delta\mu_O$ (h).

To compare the stability of the Si- and C-terminated interfaces, we evaluate the absolute interface energies as a function of both the oxygen and silicon chemical potentials, as described in Equation 3. For each value of the oxygen chemical potential $\mu_O$, we identify the most stable interface configuration within the corresponding set of structures (Si- or C-terminated) and use it to evaluate the absolute interface energy (Figure 6).

C-terminated interfaces are less stable than the best Si-terminated ones over the entire $\mu_O$ and $\mu_{Si}$ ranges. This outcome implies that Si-terminated 3C-SiC offers not only a more favorable geometric substrate but also allows oxidation to act as a stabilizing mechanism, efficiently saturating all dangling bonds and lowering the interface energy. Conversely, the C-termination neither benefits as much from oxygen incorporation nor offers a structurally smooth interface. This is important in considering as less-ordered interfaces can lead to the formation of electronic traps and ultimately degrade the performance of the heterostructure. It is also important to note that our investigation into the C-terminated interfaces is limited to specific cases; thus, it should not be taken as fully satisfying, and other structures may lead to lower interface energies.

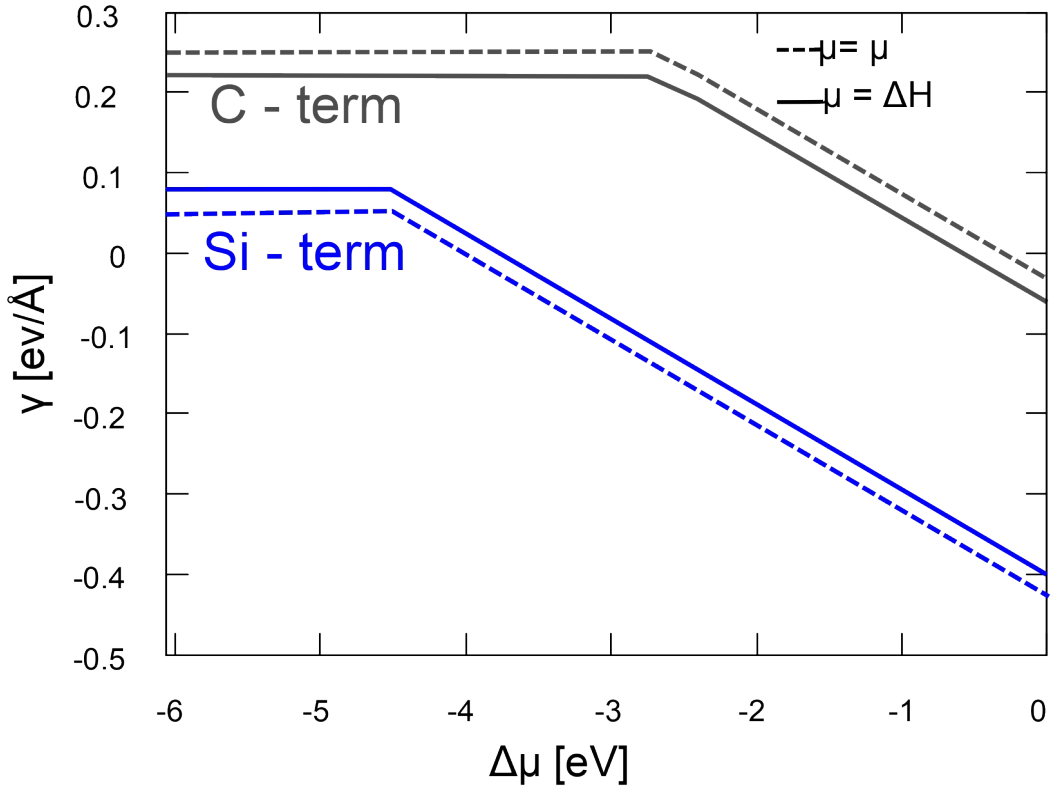

**Figure 6** - Interface energy ($\gamma_{int}$) of the most stable Si-terminated (blue) and C-terminated (gray) configurations as a function of the oxygen chemical potential $\Delta\mu_{ox}$ at fixed silicon chemical potential $\mu_{Si}$ values.

A broader perspective on our results can be obtained by comparing the interface energies computed for β-$Ga_2O_3$ on 3C-SiC with those reported for $Ga_2O_3$ interfaces on sapphire ($Al_2O_3$) [19], which represents, at present, the best substrate for $Ga_2O_3$ heteroepitaxy, thanks to a common interfacial oxygen network that joins the two materials. For this comparison, it is necessary to consider the interface energy values, which do not include an oxidation contribution. Thus, for the comparison, we used the Si-terminated interface with only stoichiometric oxygen, which is the most stable structure under Ga-rich conditions. Its interface energy values are in the range of 52-80 meV/Å$^2$, depending on $\mu_{Si}$. We reported in a previous work that the interface energy for β-$Ga_2O_3$ / α-$Al_2O_3$ heterostructure is 53 meV/Å$^2$. Thus, the two interface energies are of the same order of magnitude and we can conclude that 3C-SiC could be a promising candidate for β-$Ga_2O_3$ heteroepitaxy.

## 4) CONCLUSIONS

Our first-principles investigation provides detailed insights into the interface stability of β-$Ga_2O_3$ (100) on oxidized 3C-SiC (001) substrates, considering both Si- and C-terminated surface configurations. Our systematic analysis of interface energetics reveals several key findings with important implications for heteroepitaxial growth of $Ga_2O_3$ on Si.

The calculations demonstrate that surface oxidation of 3C-SiC substrates significantly affects interface adhesion and stability. The formation of oxidezed layers at the SiC surface prior to β-$Ga_2O_3$ deposition creates a chemically compatible interface that facilitates epitaxial growth by establishing a continuous oxygen network between substrate and film. Our results indicate that specific oxidation states and terminations exhibit markedly different adhesion energies, suggesting that careful control of pre-deposition surface preparation is essential for achieving high-quality heterostructures.

The comparison between Si- and C-terminated interfaces shows distinct thermodynamic preferences, with implications for growth orientation and domain formation. These findings provide guidance for experimental optimization of growth conditions, including substrate preparation protocols and deposition parameters. The relatively favorable interface energies calculated for certain configurations, as compared to the most convenient interface energies of $Ga_2O_3$ on conventional c-sapphire substrate, support the viability of β-$Ga_2O_3$/3C-SiC heterostructures for thermal management applications, addressing the primary limitation of poor thermal conductivity of β-$Ga_2O_3$.

## ACKNOWLEDGEMENTS

We acknowledge the CINECA consortium under the ISCRA initiative for the availability of high-performance computing resources and support.

## DATASET

The dataset of the optimized structures discussed in this work can be found at the following repository: https://doi.org/10.5281/zenodo.18669768 .

# SUPPLEMENTARY INFORMATION

*Misfit calculations*

The lattice mismatch between the most common surface cells of the α-, β-, and κ-$Ga_2O_3$ polymorphs, described in Ref.[18], against different reconstructions of 3C-SiC (001) surfaces was calculated for rectangular surface lattices as:

$$f_i = \frac{N_i \cdot a^{film}_i - M_i \cdot a^{substrate}_i}{N_i a^{film}_i} \quad ; \quad i = x, y \, ; 1 < M, N < 8$$

where $a_i^{film}$ and $a_i^{substrate}$ are the lattice parameter of the film and the substrate, respectively, along the selected i crystallographic direction, for supercells of different size, specified through the *M and N* integer numbers, which define the epitaxial matrix.

**Table S1** – Lattice parameters mismatch values $f_x/f_y$ (reported as %, white cells) and the corresponding epitaxial matrices *M, N* (gray cells) for the different combinations of surfaces and phases of $Ga_2O_3$ matching the in-plane lattice of different reconstructions of the 3C-SiC (001) surface. Negative *f* values indicate the film cell parameter is smaller than the substrate one, vice versa for positive values. The label "R90" in the epitaxial matrix indicates that the film cell is rotated by 90° with respect to the given orientation of the substrate. Below the in-plane directions for each surface are reported.

|  | β (-201) | β (100) | β (001) | κ (100) | κ (010) | κ (001) | α (001) |
|---|---|---|---|---|---|---|---|
| 001 (1x1) | 4,7 / -3,3 | 4,7 / 0,0 | 0,5 / 7,7 | -0,5 / 6,2 | -0,4 / 6,2 | -0,5 / -0,4 | -1,7 / -0,7 |
|  | 3x2 \| 2x7 | 3x3 \| 2x4 | 3x2 \| 2x5 | 6x1 \| 7x2 | 1x1 \| 2x2 | 6x1 \| 7x2 | 6x1 \| 7x2 |
| 001 p(1x1) | -1,1 / -4,4 | -1,1 / -6,0 | -5,5 / -4,4 | -1,5 / 0,5 | -6,5 / 0,5 | -1,5 / -6,5 | -2,7 / -6,8 |
|  | 1x1 \| 1x5 | 1x1 \| 1x2 | 1x1 \| 1x4 | 3x1 \| 5x3 | 1x1 \| 3x3 | 3x1 \| 5x3 | 3x1 \| 5x3 |
| 001 p(2x1) | -1,1 / -4,4 | -6,0 / -1,1 | -4,4 / -5,5 | 0,5 / -1,5 | -0,6 / 0,5 | -1,5 / -6,5 | -2,7 / -6,8 |
|  | 2x1 \| 1x5 | 1x1 \| 1x1 | 1x1 \| 2x1 | 2x3 \| 3x5 | 5x1 \| 7x3 | 6x1 \| 5x3 | 6x1 \| 5x3 |

|       |        | R 90°       | R 90°       | R 90°       |             |             |             |             |
|-------|--------|-------------|-------------|-------------|-------------|-------------|-------------|-------------|
| 001   |        | -1,1 / 2,5  | -1,1 / -6,0 | -5,5 / -4,4 | -1,5 / 0,5  | 0,6 / 0,5   | -1,5 / -0,6 | 1,4 / 0,4   |
| c(2x2)|        | 2x3 \| 1x7  | 2x1 \| 1x1  | 2x1 \| 1x2  | 6x2 \| 5x3  | 5x2 \| 7x3  | 6x5 \| 5x7  | 5x5 \| 4x7  |
| 110   |        | 4,5 / -4,4  | 4,7 / -6,0  | 0,5 / -4,4  | 5,4 / -3,6  | -0,4 / 0,5  | -0,4 / 0,5  | 0,5 / -6,5  |
|       |        | 3x1 \| 2x5  | 3x1 \| 2x2  | 3x1 \| 2x4  | 3x1 \| 2x5  | 6x1 \| 7x3  | 1x1 \| 2x3  | 6x1 \| 7x3  |

| | 3C-SiC | | | | |
|---|---|---|---|---|---|
|   | 001 (1x1) | 001 p(1x1) | 001 p(2x1) | 001 c(2x2) | 110 |
| x | [100]     | [110]      | [110]      | [110]      | [001] |
| y | [010]     | [1-10]     | [1-10]     | [1-10]     | [1-10] |

| | $Ga_2O_3$ | | | | | | |
|---|---|---|---|---|---|---|---|
|   | β (-201) | β (100) | β (001) | κ (100) | κ (010) | κ (001) | α (001) |
| x | [010]    | [001]   | [100]   | [010]   | [100]   | [100]   | [100]   |
| y | [102]    | [010]   | [010]   | [001]   | [001]   | [010]   | [010]   |

*Oxidation of the p(2x1) Si-terminated 3C-SiC (001) surface*

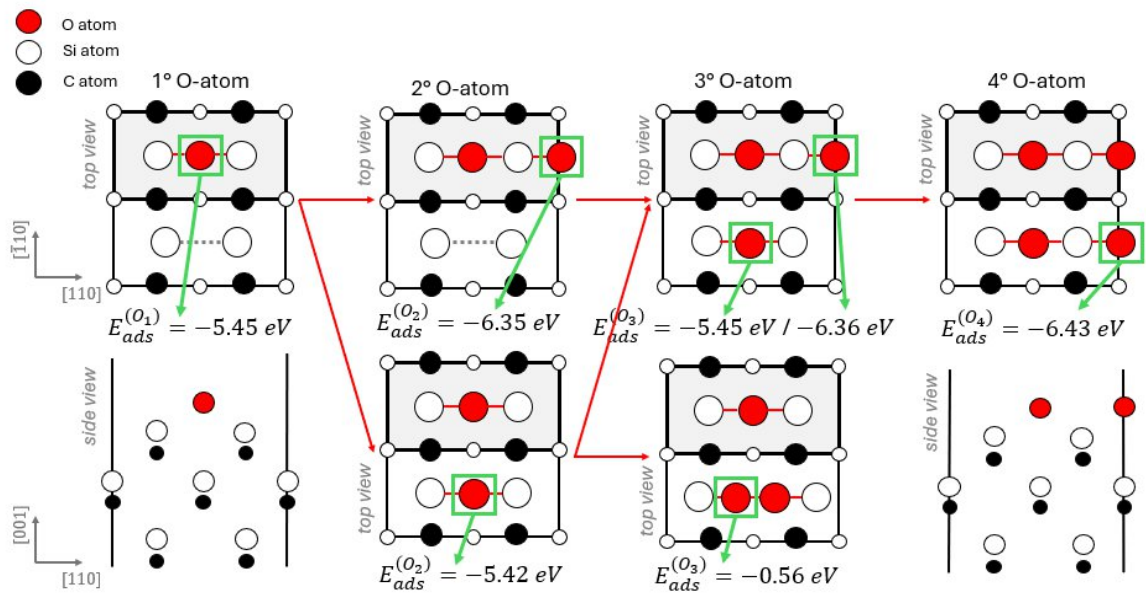

**Figure S1** – optimized structure and adsorption energy of different surface-oxidation configurations of the p(2x1) Si-terminated 3C-SiC (001) surface. For each structure, the reported adsorption energies are referred to the additional oxygen adsorbed on the surface, marked by a green square.

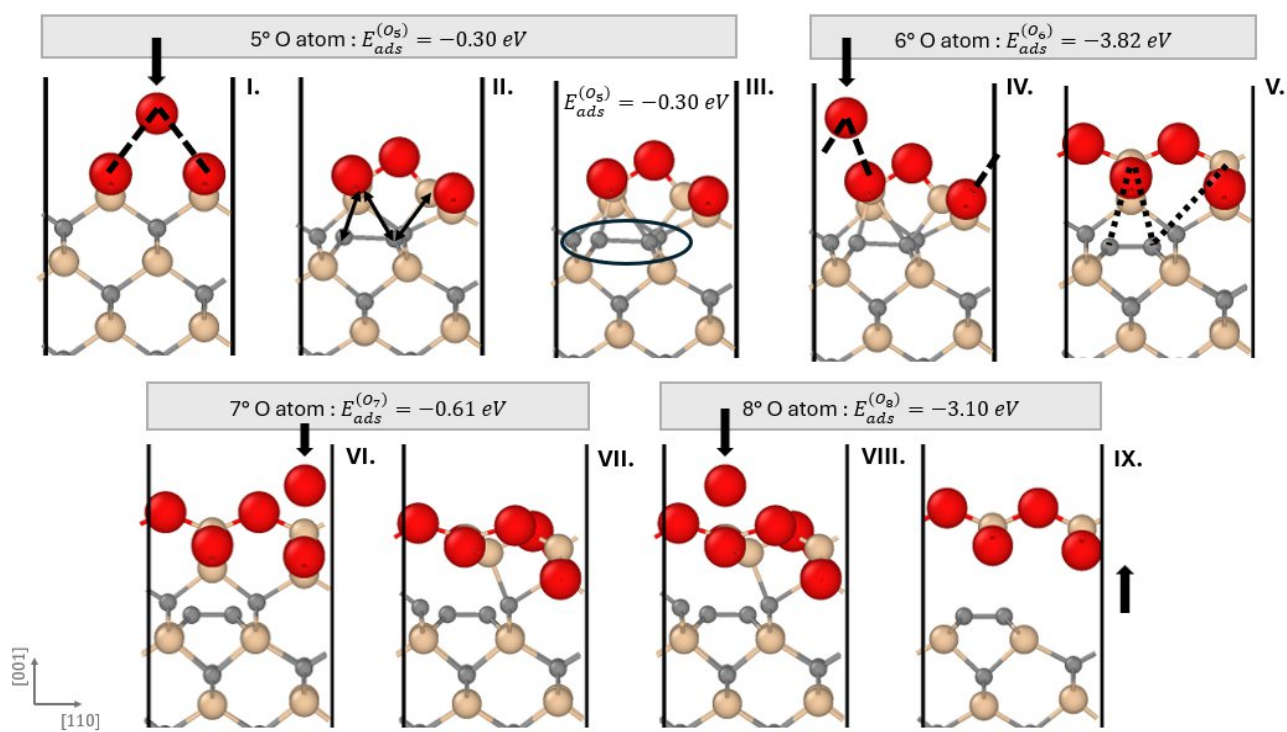

**Figure S2** –Schematic representation of the optimization steps of on a 3C-SiC (001) surface, terminated by a complete monolayer of O adatoms with the addition of 5 to 8 oxygen atoms (one at a time, marked by black arrows). A fifth oxygen atom is added to the full coverage configuration and it strongly interacts with the surface Si atoms (panel I); consequently the superficial Si atoms moves closer and elongate their bonds with the underlying C atoms (panel II), so the carbon atoms also shift and form a C–C dimer (panel III); upon the addition a further oxygen atom, which again interacts with the Si atoms (panel IV), the Si–C bonds are finally broken (panel V). The same mechanism can be observed when adding a seventh O atom (panel VI) and an eighth one (panel VIII), leading to the detachment of the $SiO_2$ layer from the surface (panel IX).

## Oxidation of the bridge $C_2$ reconstructed C-terminated 3C-SiC (001) surface

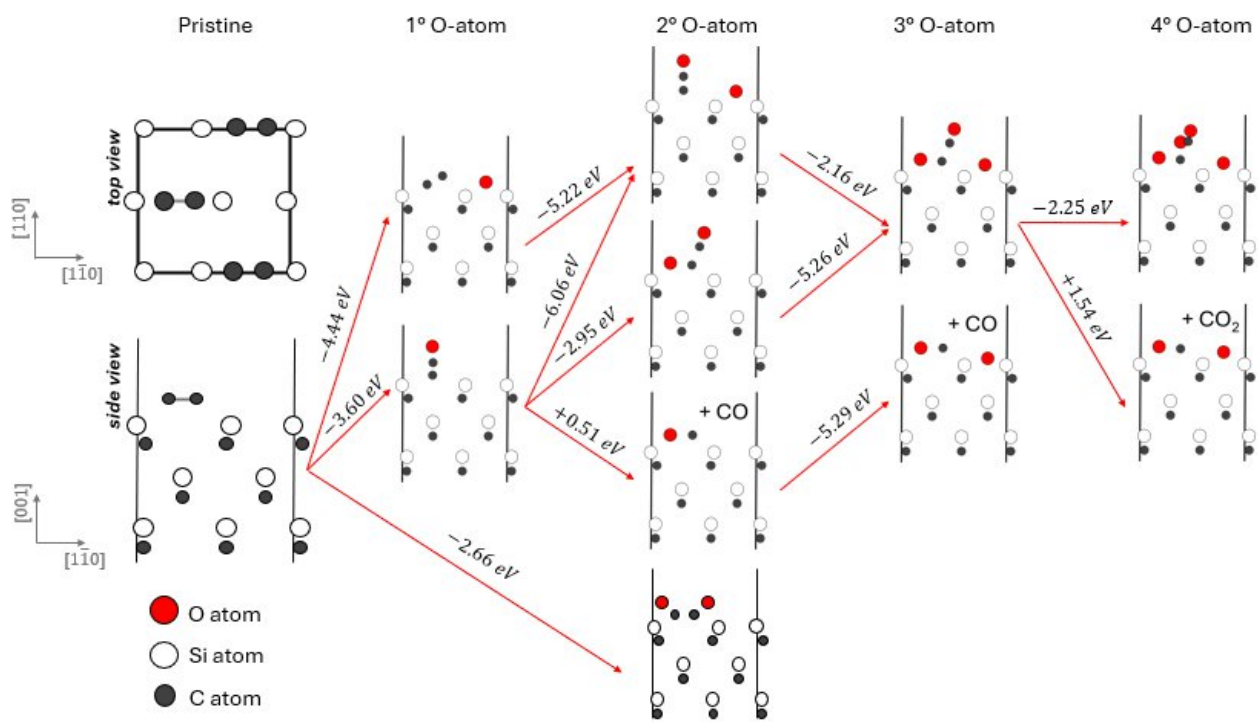

**Figure S3** - Schematic overview of the oxidized configurations of the bridge $C_2$ c(2x2) C-terminated (001) surface that we optimized, with up to four oxygen atoms per $C_2$ group. The progressive adsorption energies, e.g. the energy difference due to the adsorption of the marked adatom, are reported above the red arrows. On the left, the structure of the pristine, non-oxidized, reconstructed surface.

Figure S4 illustrates the stepwise oxidation of the $C_2$ bridging model of c(2×2) C-terminated 3C-SiC (001) surface by progressively adding oxygen atoms to a single $C_2$ group (the same behaviour was found for both equivalent groups). Each column corresponds to a different oxygen coverage, from the pristine surface up to four oxygen atoms, and shows the relaxed structures in side view. The red arrows indicate the oxidation pathways connecting configurations with successive oxygen content, while the values reported next to each arrow represent the adsorption energy of one oxygen atom associated with the corresponding step. Negative values denote energetically favored adsorption processes, whereas positive values indicate unfavored ones. Multiple arrows are shown whenever different final configurations can be reached depending on the initial adsorption site.

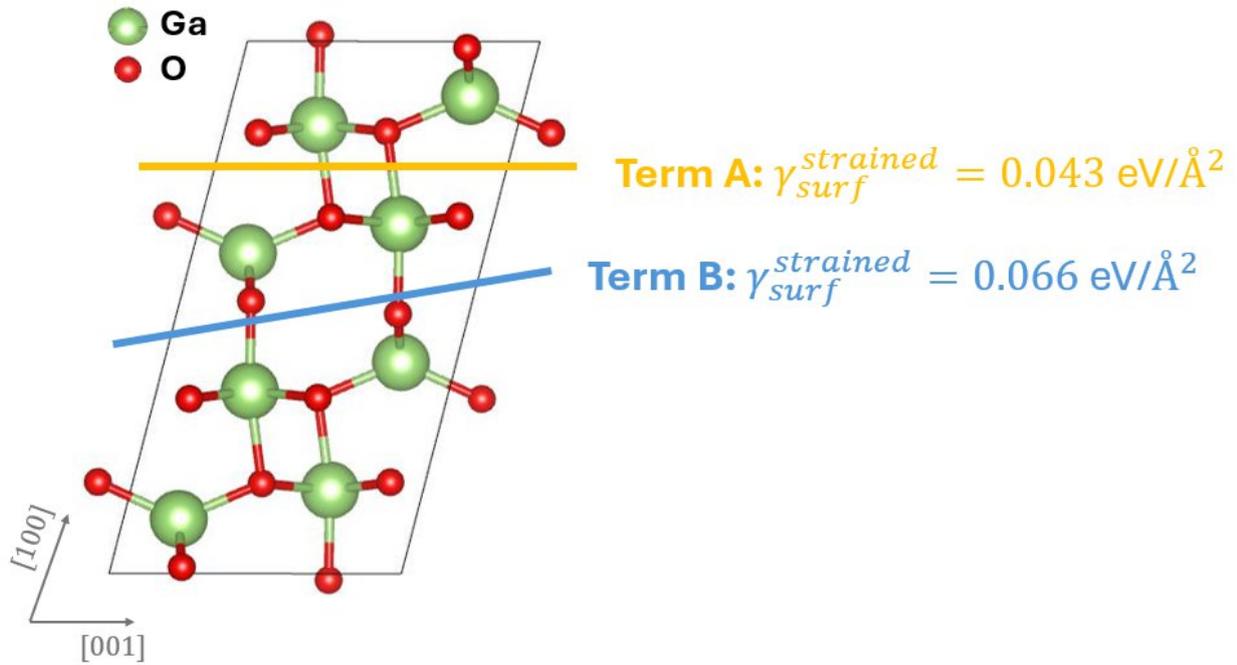

**Figure S4** - Side view of the β-Ga$_2$O$_3$ (100) surface illustrating the two inequivalent terminations, labelled A and B, and the surface energy values that we calculated for the strained surface after matching its lattice parameters to those of the 3C-SiC (001) substrate.

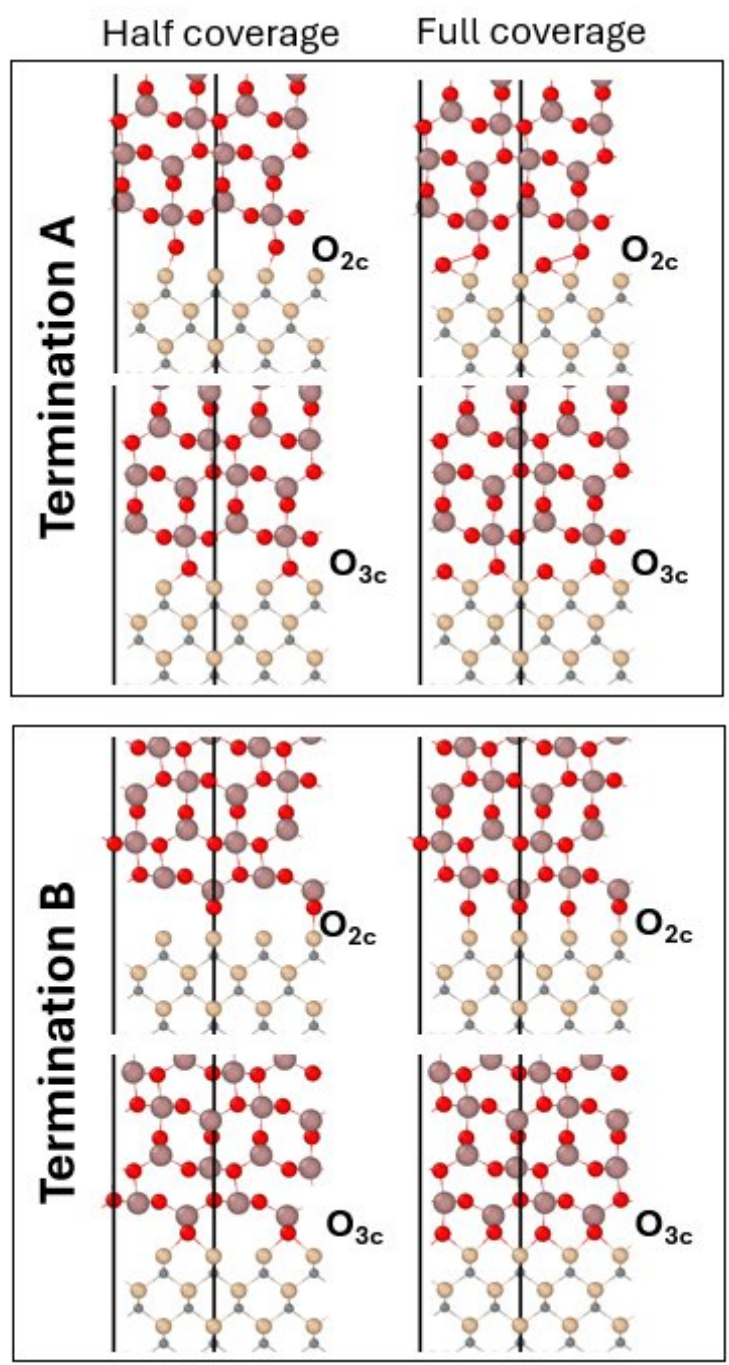

**Figure S5** - Schematic representation of the initial structures of the eight configurations of *β*-Ga$_2$O$_3$ (100)/Si-terminated 3C-SiC (001) interfaces investigated in this study, differing for the *β*-Ga$_2$O$_3$ termination (A or B), the oxidation coverage (half or full), and the initial oxygen coordination (2 or 3).

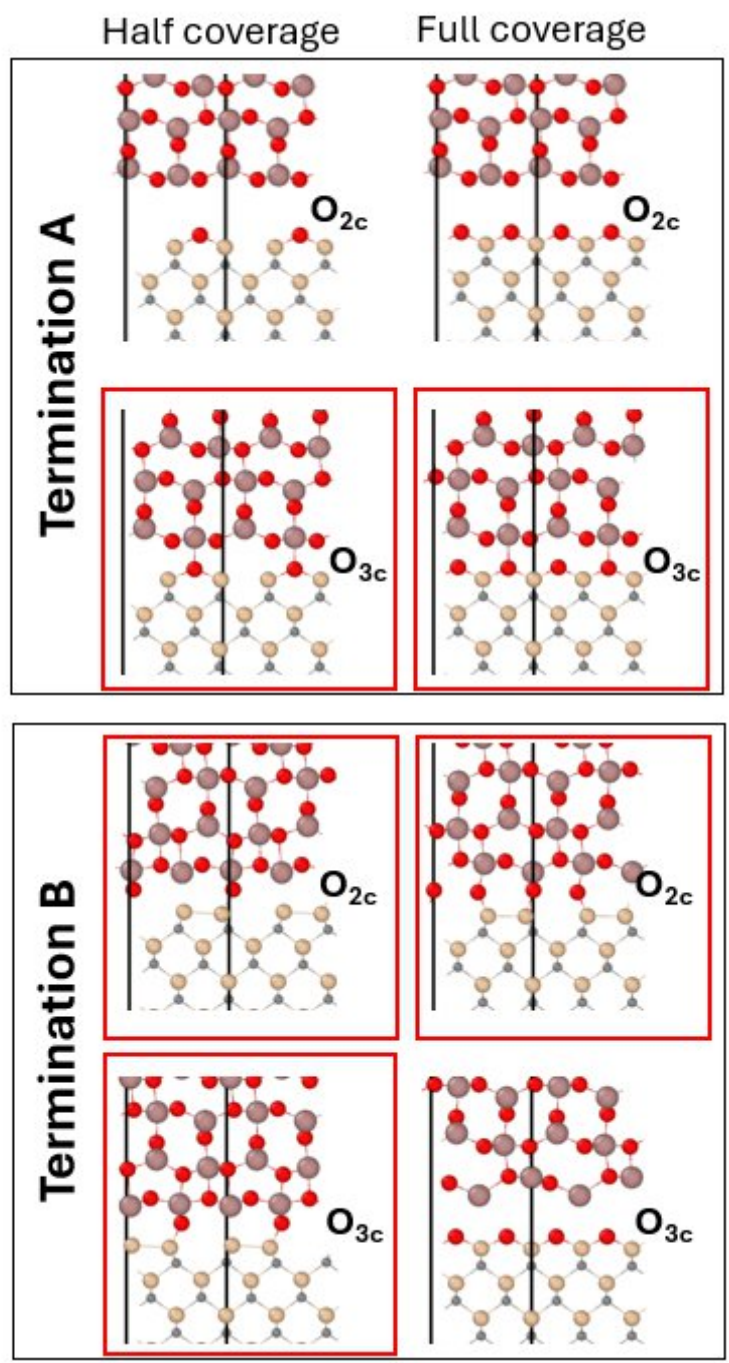

**Figure S6** - Schematic representation of the optimized structures of the eight configurations of *β*-Ga$_2$O$_3$ (100)/Si-terminated 3C-SiC (001) interfaces investigated in this study, differing for the *β*-Ga$_2$O$_3$ termination (A or B), the oxidation coverage (half or full), and the initial oxygen coordination (2 or 3). The ones which result stable and do not move apart are boxed in red.